\documentclass[12pt,a4paper]{article}
\usepackage[utf8x]{inputenc}
\usepackage{jheppub}
\usepackage{rotating}
\usepackage{bm}        
\usepackage{amssymb}   
\usepackage{dsfont}
\usepackage{tikz-feynman}
\usepackage{mathrsfs}
\usepackage{soul}
\usepackage{caption}
\usepackage{subcaption}

\def\be{\begin{equation}}
	\def\ee{\end{equation}}
\def\beal{\begin{equation}\begin{aligned}}
		\def\eeal{\end{aligned}\end{equation}}

\def\bra#1{\langle #1|}
\def\ket#1{|#1 \rangle}
\def\braket#1{\langle #1 \rangle}

\renewcommand{\[}{\begin{equation}\begin{aligned}}
\renewcommand{\]}{\end{aligned}\end{equation}}
\newcommand{\scri}{{\mathscr I}}
\renewcommand{\vec}[1]{\ensuremath{\boldsymbol{#1}}}
\newcommand{\vac}{\Omega}

\newcommand{\sfk}{\mathsf{k}}

\newcommand{\dd}{\mathrm{d}}

\newcommand{\rvaidya}{$\sqrt{\textrm{Vaidya}}$ }

\title{Hawking Radiation meets the Double Copy}
\author[1]{Rafael Aoude,}
\emailAdd{rafael.aoude@ed.ac.uk}
\author[1]{Donal O'Connell,}
\emailAdd{donal@ed.ac.uk}
\author[2,3]{Matteo Sergola,}
\emailAdd{msergola@physics.ucla.edu}
\author[4]{Chris D. White}
\emailAdd{christopher.white@qmul.ac.uk}

\affiliation[1]{Higgs Centre for Theoretical Physics,
School of Physics and Astronomy, \\
The University of Edinburgh, Edinburgh EH9 3JZ, Scotland, UK}
\affiliation[2]{
Institut de Physique Théorique, CEA, CNRS, \\Université Paris-Saclay, F–91191 Gif-sur-Yvette cedex, France}
\affiliation[3]{Mani L. Bhaumik Institute for Theoretical Physics,
University of California at Los Angeles, Los Angeles, CA 90095, USA}
\affiliation[4]{Centre for Theoretical Physics, School of Physical and Chemical Sciences, Queen Mary University
of London, 327 Mile End Road, London E1 4NS, United Kingdom}

\abstract{We describe an electromagnetic system which is related to black hole production with Hawking radiation through the double copy. We consider the scattering of a massless scalar particle through a collapsing electromagnetic background -- the single copy of Vaidya -- and identify the Feynman diagrams that exponentiate in the geometric-optics limit. The Bogoliubov coefficients obtained from the diagrammatic approach are reproduced by a semiclassical ray-tracing computation of null rays in this same background. We discuss the thermodynamic interpretation of the resulting number distribution in light of the double copy.}

\begin{document}
\maketitle

\section{Introduction}
\label{sec:intro}

The study of scattering amplitudes in gauge and gravity theories
continues to attract attention. Two major topics in recent years have
been the use of amplitudes to reveal intriguing connections between
different (quantum) field theories; also, their ability to shed new
light on questions in (semi-)classical physics, particularly those
relating to black holes. Motivated by this, ref.~\cite{Aoude:2024sve}
considered the well-known phenomenon of Hawking
radiation~\cite{Hawking:1974rv,Hawking:1975vcx} (see
e.g. refs.~\cite{Harlow:2014yka,Page:2004xp} for extensive reviews),
addressing the problem using amplitude methods developed in the
context of gravitational wave physics~\cite{Kosower:2018adc} (see also
refs.~\cite{Goldberger:2020geb,Goldberger:2020wbx,Kim:2020dif,Ilderton:2023ifn,Gaddam:2021zka,Gaddam:2020mwe,Ferreira:2020whz,Melville:2023kgd,Aoude:2023fdm,Chen:2023qzo,Vidal:2024inh,Aoki:2025ihc,Ilderton:2025umd,Clark:2025tqi}
for related ideas). Motivated by Hawking's original calculation, the
authors considered the metric corresponding to a collapsing shell of
null dust, and a one-to-one scattering amplitude of a scalar particle
in this background. By resumming a set of all possible Feynman diagrams
involving interactions with the background field, the authors could
relate the amplitude to the thermal spectrum of emitted particles. One
way to understand this is through the known interpretation of Hawking
radiation as pair creation of particles near the black hole horizon,
followed by quantum tunnelling of one of the
particles~\cite{Parikh:1999mf}. The pair creation amplitude is related
to the Hawking (one-to-one) amplitude by crossing.

Scattering amplitudes in gravity theories are related to those in
gauge theory by the double
copy~\cite{Bern:2008qj,Bern:2010ue,Bern:2010yg}, itself inspired by
previous work in string theory~\cite{Kawai:1985xq}. The double copy
has also been extended to classical
solutions~\cite{Monteiro:2014cda,Luna:2015paa,Ridgway:2015fdl,Bahjat-Abbas:2017htu,Carrillo-Gonzalez:2017iyj,CarrilloGonzalez:2019gof,Bah:2019sda,Alkac:2021seh,Alkac:2022tvc,Luna:2018dpt,Sabharwal:2019ngs,Alawadhi:2020jrv,Godazgar:2020zbv,White:2020sfn,Chacon:2020fmr,Chacon:2021wbr,Chacon:2021hfe,Chacon:2021lox,Dempsey:2022sls,Emond:2021lfy,Easson:2022zoh,Chawla:2022ogv,Han:2022mze,Armstrong-Williams:2022apo,Han:2022ubu,Elor:2020nqe,Farnsworth:2021wvs,Anastasiou:2014qba,LopesCardoso:2018xes,Anastasiou:2018rdx,Luna:2020adi,Borsten:2020xbt,Borsten:2020zgj,Goldberger:2017frp,Goldberger:2017vcg,Goldberger:2017ogt,Goldberger:2019xef,Goldberger:2016iau,Prabhu:2020avf,Luna:2016hge,Luna:2017dtq,Cheung:2016prv,Cheung:2021zvb,Cheung:2022vnd,Cheung:2022mix,Chawla:2024mse,Keeler:2024bdt,Chawla:2023bsu,Easson:2020esh,Armstrong-Williams:2024bog,Armstrong-Williams:2023ssz,Farnsworth:2023mff,Emond:2025nxa,Moynihan:2025vcs,Ilderton:2025gug,Kent:2024mow,Easson:2023dbk,Easson:2022zoh,Keeler:2020rcv}
(see
e.g. refs.~\cite{Borsten:2020bgv,Bern:2019prr,Adamo:2022dcm,Bern:2022wqg,White:2021gvv,White:2024pve}
for recent reviews), where a canonical formalism is that of the {\it
  Kerr-Schild double copy} of ref.~\cite{Monteiro:2014cda}. Given that
the analysis of ref.~\cite{Aoude:2024sve} adopted a Kerr-Schild form
for the gravitational background field, the authors already posed the
question of whether a single copy of their calculation exists, and can
be interpreted. The aim of this paper is to carry out this
investigation, and there are a number of motivations for doing
so. Firstly, despite a broad literature on the double copy, its
ultimate origin and scope remain somewhat mysterious. Concrete
examples of how known physics in either gauge or gravity theories can
be directly related is highly sought after, providing much-needed
physical intuition of how the double copy operates. Indeed, such
insights may be useful for the more general web of QFTs that are now
known to be related by double-copy-like correspondences. Secondly, a
single copy of Hawking radiation may be valuable for understanding
further quantum properties of black holes. The study of semi-classical
gravity remains an ongoing research area, with many open questions,
including the precise nature of potential microstates that can lead to
the known black hole entropy. Knowing that one can obtain Hawking
radiation from a double copy opens the door for finding further black
hole properties, by recycling results from a simpler gauge theory. For
completeness, we note that gauge theory analogues of Hawking radiation
have an established
history~\cite{Parikh:1999mf,Wondrak:2023zdi,Wondrak:2023hcz,Stephens:1989fb,Parentani:1991tx,Kim:2011fs,Kim:2012wg,Srinivasan:1998ty}. Our
work is complementary to these, in that the nature of the analogue is
determined by the particular relationship of the single copy, which
offers alternative possibilities for extending our approach to obtain
additional insights.

The structure of our paper is as follows. In
section~\ref{sec:background}, we find and interpret the single copy of
the classical (Vaidya) spacetime used in ref.~\cite{Aoude:2024sve} to
derive Hawking radiation from an appropriate scattering amplitude. In
sec.~\ref{sec:BogQFT}, we find the gauge theory analogue of this
amplitude, and spell out its relation to the Bogoliubov coefficients
needed to examine particle production properties. We cross-check our results in
sec.~\ref{sec:Bogray} by comparing with a ray-tracing calculation, as
described in ref.~\cite{Aoude:2024sve}. In sec.~\ref{sec:number}, we
examine the number spectrum of emitted particles from the single-copy
Vaidya solution. This shows crucial differences from the gravitational
case, and we will speculate regarding the physical interpretation of
our results. Finally, we discuss our results and conclude in
sec~\ref{sec:conclude}. Appendix \ref{appendix} proves energy-momentum conservation for the non-static background current.

\subsubsection*{Note added} In the final stages of this project  we learned about parallel research in references \cite{Ilderton:2025,Carrasco:2025bgu} which contain some overlap with our work and are to appear in forthcoming articles. We thank the authors for cooperating with us in the submission of our work and for sharing advance copies of their drafts. We have also checked that our results are compatible by comparing our eikonal with the wavefunctions computed in \cite{Ilderton:2025}.  

\section{The classical background}
\label{sec:background}

In this section, we begin our detailed discussion of the single-copy of Hawking radiation. 
We will follow the basic logic of Hawking's original paper, computing Bogoliubov coefficients $A$ and $B$, which describe the dynamics of a massless scalar field in a time-dependent background, and extracting the statistical number distribution from $B$.
Let us start by developing an understanding of the relevant background.

Hawking considered a situation involving some matter collapsing to form a black hole.
A very simple example of a collapse background is given by the Vaidya metric~\cite{Vaidya:1966zza,Stephani_Kramer_MacCallum_Hoenselaers_Herlt_2003}
\[
g_{\mu\nu} = \eta_{\mu\nu} - \frac{2 GM}{r} \Theta(t+r) \sfk_\mu \sfk_\nu \,,
\]
where $M$ is the mass of the (future) black hole, while $\sfk \cdot \dd x = \dd(t+r)$. 
This is a Kerr-Schild metric: exact although it is linear in Newton's constant $G$.
More precisely, Vaidya metrics are a class of spacetimes involving time-dependent mass functions, but for our purposes the choice $M \Theta(t+r)$ is particularly convenient.
The Heaviside theta function corresponds to a black hole forming suddenly on the time scales of interest.
Physically, this metric describes a black hole formed from a thin shell of infalling light-like radiation ---  therefore the spacetime involves a specific stress-energy tensor describing radiation in the past.

Since the metric admits a Kerr-Schild decomposition, it is straightforward to write down a possible electromagnetic\footnote{Strictly speaking, the single copy of a gravity solution is a non-abelian gauge field. However, as discussed in the original Kerr-Schild double copy paper~\cite{Monteiro:2014cda}, this field has the form $A_\mu^a=c^a A_\mu$, where $c^a$ is a constant colour vector. This has the effect of linearising the Yang-Mills equations, so that one may choose to ignore the colour dependence.} single-copy using the usual rules of the Kerr-Schild double copy \cite{Monteiro:2014cda}: 
we strip off one Kerr-Schild vector and replace momentum by charge. In this case we also have to deal with the mass and its time dependence. It seems straightforward to extend the charges replacement of the static case as follows
\begin{align}
\label{eq:SqrtVaidya}
2GM(t+r)\to \frac{Q_{\rm B}(t+r)}{4\pi},
\end{align}
where $Q_{\rm B}(t+r)= Q_{\rm B}\Theta(t+r)$ with constant charge $Q_{\rm B}$ (``B" is for background). Extending the time dependence to the charge in this way is also consistent with the Bonnor-Vaidya metric \cite{Bonnor:1970zz} which extends the charged Reissner–Nordström solution of GR.

The result is an electromagnetic potential
\[\label{eq:potential}
A_\mu &= \frac{Q_{\rm B}}{4\pi r} \Theta(t+r)\, \sfk_\mu \,, \\ 
\Rightarrow A &= \frac{Q_{\rm B}}{4\pi r} \Theta(t+r)\, \dd (t+r) \,. 
\]
We will refer to this as the $\sqrt{\textrm{Vaidya}}$ background.

To interpret this potential, first note that the field strength two-form is
\[
\label{eq:vaidyafield}
F = \dd A = \frac{Q_{\rm B}}{4\pi r^2} \Theta(t+r)  \, \dd(t+r) \wedge \dd r \,.
\]
The field strength vanishes when $t+r <0$,
but for $t+r >0$ the field strength is just that of the Coulomb field of a static charge $Q_{\rm B}$ at the origin.
Therefore, on a narrow shell at $t+r = 0$ there must be some infalling charge distribution (of total charge $Q_{\rm B}$).
When this charge distribution reaches the origin $r=0$ (at time $t=0$) it binds into a total point charge $Q_{\rm B}$. 
Obviously some exterior force is required to bind all this charge together in some small spacetime region; but this is physically acceptable in electromagnetism. 
Clearly, the situation is much less natural in electromagnetism than in gravity.

As in the gravitational case, the \rvaidya background involves a source: the current density of the infalling charges. 
Using the Maxwell equation, a short computation shows that
\[\label{current}
j_\mu = \frac{Q_{\rm B}}{4\pi r^2} \delta(t+r) \, \sfk_\mu \,.
\]
This describes a (very) thin shell of infalling massless charge.
The distributional support is a long-wavelength simplification; physically, the shell of infalling radiation must of course have a finite size.
Here we are working in the approximation that this size is negligible.
We are also assuming that the mass of the charges that make up the distribution can be completely neglected. In this section, we have seen that one may write a consistent single copy of the Vaidya spacetime, that solves the Maxwell equations with a readily interpretable source current. This is itself an interesting result, given that the original Kerr-Schild double copy was formally derived only for static solutions (although see refs.~\cite{Carrillo-Gonzalez:2017iyj}, \cite{Kent:2025pvu} for extensions). Our result shares the common feature that taking the single copy amounts to simply replacing mass by charge. In Appendix \ref{appendix} we will also see how the total energy-momentum is conserved thanks to \eqref{current}. 

\section{Bogoliubov coefficients from quantum field theory}
\label{sec:BogQFT}

Using quantum field theory and the methods of scattering amplitudes, in this section we obtain the one-to-one amplitude with a collapsing source of electromagnetic charge.

\subsection{Amplitude calculation and its eikonal resummation}
\label{sec:resum}

As discussed above, the electromagnetic background potential is simply \eqref{eq:potential}, which we report again for convenience here
\begin{align}
\label{eq:Potential2}
\quad A^\mu(x)  =\frac{Q_\text{B}(t+r)}{4\pi r}\sfk^\mu.
\end{align}
where $\sfk_\mu$ is a null vector
\begin{align}
    \sfk_\mu(x) = \left(1,\frac{{\bm x}}{r}\right),
    \qquad
    \sfk_\mu(x)\sfk^\mu(x) =0.
\end{align}
Now we follow the procedure outlined in Section 3 of~\cite{Aoude:2024sve}: we scatter a massless particle with charge $Q$ off the time-dependent background. 
Note that the massless choice is purely a convenience here. We view this as a first order approximation to a probe which is much lighter than the $\sqrt{\text{Vaidya}}$ source. Following references~\cite{Kosower:2018adc,Cristofoli:2021jas}, we specify the initial KMOC state of a charged massless particle which we take to be spherically symmetric:
\begin{align}\label{state}
    \ket{\psi} \equiv \int\! 
    \dd\Phi(p)\varphi(p)|p\rangle 
     = \int\! 
    \dd\Phi(p)\ket{p}\int \! \dd v \, e^{iEv}
    {\varphi}(v) \,,
\end{align}
where 
\[
\dd \Phi(p) \equiv \hat{\dd}^4 p \, \hat{\delta}(p^2) \Theta(p^0),
\]
is the on-shell phase space measure, and 
\begin{equation}
    \ket{p}= a^\dagger (p)\ket{\vac} \,,
\end{equation}
is a momentum eigenstate generated by acting on the vacuum $\ket{\vac}$ of the scalar theory in the  $\sqrt{\text{Vaidya }}$ potential. 
The function 
$\varphi(p)$ represents a momentum-space wavefunction which, as in ref.~\cite{Aoude:2024sve}, we may Fourier transform to a position-space wavefunction $\varphi(v)$. The latter depends only upon the single variable $v$ owing to spherical symmetry and the fact that $E=|\vec p|$ for a massless state. Then the wavefunction $\varphi(p)$ depends only on the real variable $E$, the energy of the massless state: $\varphi(p)=\varphi(E)$. The role of $\varphi(v)$ is to construct a suitable wavepacket for the particle that scatters on the electromagnetic background $A_\mu(x)$. Given that we are interested in particles that start inside the infalling spherical shell of charge, we will take $\varphi(v)$ to have support only for $v<0$ in what follows. 
Finally, to make contact with the scattering amplitudes literature we introduce a \emph{timelike} impact parameter  
\begin{equation}
    b^\mu(v)=(v,\vec{0}),
\end{equation}
so that we can covariantise the exponential as $e^{iEv}=e^{ib\cdot p}$.

At this point we are ready to determine the  time evolution of the state \eqref{state} using the $S$-matrix. For simplicity we also project onto a single particle state of momentum $p'$ and subtract the non-scattering contribution, thus obtaining 
\[
\label{sstate}
 \braket{p'|S-1|\psi}&= \int \dd\Phi(p) \int \! \dd v \, e^{ip\cdot b(v)}
    \varphi(v) \,\langle p'| i T|p\rangle 
    \\&=
    \int \dd v \,\varphi(v) e^{i p'\cdot b(v)}\int \hat{\dd}^4 q\, \hat{\delta}(2 p'\cdot q+q^2)i\mathcal{A} (p'-q\to p')e^{-iq\cdot b(v)}.
\]
Note that above we have changed the integration variable from $p$ to $q \equiv p'-p$, this is useful in view of the geometric-optics approximation that we now discuss. 

In our setup we consider a massless scalar particle with wavelength $\lambda \sim \hbar / E = \hbar / |\mathbf{p}|$ much shorter than any other length at play, in particular $ \lambda\ll|\mathbf{b}|$ (see also the discussion in~\cite{Cristofoli:2021vyo}). Further relating the impact parameter with the momentum transfer $|\mathbf{b}| = {\hbar}/{|\mathbf{q}|}$, we define the geometric-optics limit as in  \cite{Aoude:2024sve}
\[\label{golimit}
|\mathbf{b}|\gg \lambda,
\qquad
|\mathbf{q}|\ll |\mathbf{p}| 
\qquad
\textrm{or}
\qquad
\eta \equiv |\mathbf{q}|/|\mathbf{p}| \ll 1.
\] 
Note two things. First,  because the probe is massless, $p$ can scale with $\hbar$ but we can still take the particle to be hard $p\gg q$, and use the same separation of scales as in usual Post-Minkowskian perturbations. Secondly, one can expand both in $\eta$ and the couplings $QQ_\text{B}$. Below, we will first compute the leading-$\eta$ and leading-$QQ_\text{B}$ term and then resum the leading-$\eta$ to all loop orders  $\mathcal{O}\left((QQ_\text{B})^L\right)$ through eikonalization.

Next, we model the dynamics of the system by minimally coupling the scalar field to the background. Furthermore, because of Kerr-Schild coordinates, we know that $A^2=0$. This is sometimes known as the ``Kerr-Schild" gauge \cite{Menezes:2022tcs}. Choosing this gauge makes diagrammatic interactions cubic with no contact interactions, i.e. $S_\text{int}=\mathcal{O}(A\phi^*\partial \phi)$ exactly. As we will see, this will  greatly simplify  perturbation theory. The $1\to 1$ amplitude can be obtained from standard diagrammatic methods and is represented at leading order by the momentum space diagram below. Here, the massless probe with momentum $p$ interacts with the background exchanging a photon  
\[\label{}
\begin{tikzpicture}[x=0.75pt,y=0.75pt,yscale=-1,xscale=1]

  
\tikzset {_exlviwant/.code = {\pgfsetadditionalshadetransform{ \pgftransformshift{\pgfpoint{0 bp } { 0 bp }  }  \pgftransformrotate{0 }  \pgftransformscale{2 }  }}}
\pgfdeclarehorizontalshading{_blhbr1q3h}{150bp}{rgb(0bp)=(1,1,1);
rgb(45.80357142857143bp)=(1,1,1);
rgb(54.910714285714285bp)=(0.82,0.29,0.35);
rgb(62.5bp)=(0.73,0.15,0.22);
rgb(62.5bp)=(0.95,0.56,0.6);
rgb(100bp)=(0.95,0.56,0.6)}
\tikzset{every picture/.style={line width=0.75pt}} 

\draw    (287.18,20721.69) -- (241,20713) ;
\draw [shift={(264.09,20717.35)}, rotate = 190.66] [fill={rgb, 255:red, 0; green, 0; blue, 0 }  ][line width=0.08]  [draw opacity=0] (5.36,-2.57) -- (0,0) -- (5.36,2.57) -- cycle    ;
\draw    (333,20714) -- (287.18,20721.69) ;
\draw [shift={(310.09,20717.85)}, rotate = 170.47] [fill={rgb, 255:red, 0; green, 0; blue, 0 }  ][line width=0.08]  [draw opacity=0] (5.36,-2.57) -- (0,0) -- (5.36,2.57) -- cycle    ;
\draw [color={rgb, 255:red, 0; green, 0; blue, 0 }  ,draw opacity=1 ]   (287.18,20721.69) .. controls (288.85,20723.36) and (288.85,20725.02) .. (287.18,20726.69) .. controls (285.51,20728.36) and (285.51,20730.02) .. (287.18,20731.69) .. controls (288.85,20733.36) and (288.85,20735.02) .. (287.18,20736.69) .. controls (285.51,20738.36) and (285.51,20740.02) .. (287.18,20741.69) .. controls (288.85,20743.36) and (288.85,20745.02) .. (287.18,20746.69) .. controls (285.51,20748.36) and (285.51,20750.02) .. (287.18,20751.69) .. controls (288.85,20753.36) and (288.85,20755.02) .. (287.18,20756.69) .. controls (285.51,20758.36) and (285.51,20760.02) .. (287.18,20761.69) .. controls (288.85,20763.36) and (288.85,20765.02) .. (287.18,20766.69) -- (287.18,20766.69) ;
\draw    (299,20741) -- (299,20754) ;
\draw [shift={(299,20738)}, rotate = 90] [fill={rgb, 255:red, 0; green, 0; blue, 0 }  ][line width=0.08]  [draw opacity=0] (5.36,-2.57) -- (0,0) -- (5.36,2.57) -- cycle    ;
\path  [shading=_blhbr1q3h,_exlviwant] (287.23,20757.67) .. controls (292.45,20757.7) and (296.65,20761.76) .. (296.62,20766.74) .. controls (296.59,20771.72) and (292.34,20775.74) .. (287.12,20775.71) .. controls (281.91,20775.69) and (277.7,20771.63) .. (277.73,20766.65) .. controls (277.77,20761.66) and (282.02,20757.65) .. (287.23,20757.67) -- cycle ; 
 \draw   (287.23,20757.67) .. controls (292.45,20757.7) and (296.65,20761.76) .. (296.62,20766.74) .. controls (296.59,20771.72) and (292.34,20775.74) .. (287.12,20775.71) .. controls (281.91,20775.69) and (277.7,20771.63) .. (277.73,20766.65) .. controls (277.77,20761.66) and (282.02,20757.65) .. (287.23,20757.67) -- cycle ; 

\draw (337,20706) node [anchor=north west][inner sep=0.75pt]    {$p'$};
\draw (227,20710) node [anchor=north west][inner sep=0.75pt]    {$p$};
\draw (305,20742) node [anchor=north west][inner sep=0.75pt]    {$q$};
\draw (371.67,20738.67) node [anchor=north west][inner sep=0.75pt]    {$=i\mathcal{A}_0(p\to p')=-iQ \tilde{A}^\mu(q) (2p_\mu+q_\mu).$};

\end{tikzpicture}
\]
The subscript on $\mathcal{A}_0$ indicates a tree level  interaction and $\tilde{A}^\mu(q)$ is the Fourier transform of \eqref{eq:Potential2}. The gradient on the blob is meant to graphically represents the time dependence of the EM source.

We can now begin the explicit computation of \eqref{sstate}. Retaining the leading-in-$\eta$ term in the geometric-optics limit $p\gg q$, the LO amplitude reads
\[ \label{leadinga0}
i\mathcal{A}_0(p\to p')=-2iQ \tilde{A}(q)\cdot p=\frac{-iQQ_\text{B}}{2\pi}\int \!\dd^4 x \,e^{iq\cdot x} \frac{\Theta(t+r)}{r}\sfk(x)\cdot p.
\]
We will be interested in a position space expression so it is convenient to work in impact parameter space. Fourier transforming this expression and approximating in the geometric-optics limit,
\[
\hat{\delta}(2 p'\cdot q+q^2)\approx \hat{\delta}(2 p'\cdot q),
\]
we have
\[\label{state1}
\braket{p|i{T}_{\text{tree}}|\psi}&= \frac{-iQQ_\text{B}}{2\pi}
\int\! \dd v \,\varphi(v) e^{i p\cdot b(v)}  
\int\! \dd^4 x \,\hat{\dd}^4 q\, \hat{\delta}(2 p\cdot q)
\frac{\Theta(t+r)}{r}\sfk(x)\cdot p\, e^{iq\cdot (x-b(v))},
\]
having also dropped the prime on $p'\approx p$. At this point we note that the combination of integrals above can be nicely rewritten by introducing a worldline parameter $\lambda$:
\[
\int \dd^4 x \,\hat{\dd}^4 q\, \hat{\delta}(2 p\cdot q)
\frac{\Theta(t+r)}{r}\sfk(x)\cdot p\, e^{iq\cdot (x-b(v))}=
\int \dd \lambda\left(\frac{\Theta(t+r)}{r}\sfk(x)\cdot p\right)\bigg{\rvert}_{x=b+2\lambda p} \,.
\]
This integral can be readily integrated  noting that $r=2E|\lambda|$, $t =v+2E\lambda$. We find
\[\label{eq:IRphase}
\int \dd \lambda\left(\frac{\Theta(t+r)}{r}\sfk(x)\cdot p\right)\bigg{\rvert}_{x=b+2\lambda p}=\int_{-v/(4E)}^\infty  \frac{\dd \lambda}{\lambda}=-\log(-v/\mu) +(\text{IR-phase}).
\]
Note that above we have introduced a dimensional parameter $\mu$ to make sense of the logarithm's mass dimension and restricted the integral domain with the step function $\Theta(t+r)=\Theta(v+4E\lambda)$. Furthermore, we will consistently drop the divergent --- yet \emph{constant} in $v$ ---  infrared phase that originates from the upper integration\,\footnote{See~\cite{Lippstreu:2023vvg,Lippstreu:2025jit} for more details on the treatment of infrared-finite amplitudes.}. We can do this safely since, as we will see below, this contribution resums into a $v$-independent phase that eventually cancel out in the physical spectrum.
In the end we obtain 
\[\label{treelvl}
\braket{p|i{T}_{\text{tree}}|\psi}&= 
\int \dd v \,\varphi(v) e^{i p\cdot b(v)}  \times \frac{iQQ_\text{B}}{2\pi}\log(-v/\mu).
\]

In order to get the particle spectrum, it is imperative to resum this result 
in the usual eikonal sense. Despite the different numerator, the pole structure is the same as the gravitational case~\cite{Aoude:2024sve}, and we can follow the same steps. We only need to note the momentum space expression of the leading-eikonal $L$-loop amplitude  below
\[
\begin{tikzpicture}[x=0.75pt,y=0.75pt,yscale=-1,xscale=1]

  
\tikzset {_anox4r0d0/.code = {\pgfsetadditionalshadetransform{ \pgftransformshift{\pgfpoint{0 bp } { 0 bp }  }  \pgftransformrotate{0 }  \pgftransformscale{2 }  }}}
\pgfdeclarehorizontalshading{_o1zq98a2s}{150bp}{rgb(0bp)=(1,1,1);
rgb(45.80357142857143bp)=(1,1,1);
rgb(54.910714285714285bp)=(0.82,0.29,0.35);
rgb(62.5bp)=(0.73,0.15,0.22);
rgb(62.5bp)=(0.95,0.56,0.6);
rgb(100bp)=(0.95,0.56,0.6)}

  
\tikzset {_19cp6sbk5/.code = {\pgfsetadditionalshadetransform{ \pgftransformshift{\pgfpoint{0 bp } { 0 bp }  }  \pgftransformrotate{0 }  \pgftransformscale{2 }  }}}
\pgfdeclarehorizontalshading{_ol8dd6hbu}{150bp}{rgb(0bp)=(1,1,1);
rgb(45.80357142857143bp)=(1,1,1);
rgb(54.910714285714285bp)=(0.82,0.29,0.35);
rgb(62.5bp)=(0.73,0.15,0.22);
rgb(62.5bp)=(0.95,0.56,0.6);
rgb(100bp)=(0.95,0.56,0.6)}

  
\tikzset {_qnr8f0d09/.code = {\pgfsetadditionalshadetransform{ \pgftransformshift{\pgfpoint{0 bp } { 0 bp }  }  \pgftransformrotate{0 }  \pgftransformscale{2 }  }}}
\pgfdeclarehorizontalshading{_5fkrmtb6k}{150bp}{rgb(0bp)=(1,1,1);
rgb(45.80357142857143bp)=(1,1,1);
rgb(54.910714285714285bp)=(0.82,0.29,0.35);
rgb(62.5bp)=(0.73,0.15,0.22);
rgb(62.5bp)=(0.95,0.56,0.6);
rgb(100bp)=(0.95,0.56,0.6)}

  
\tikzset {_ab148gheg/.code = {\pgfsetadditionalshadetransform{ \pgftransformshift{\pgfpoint{0 bp } { 0 bp }  }  \pgftransformrotate{0 }  \pgftransformscale{2 }  }}}
\pgfdeclarehorizontalshading{_lmmfwi4xg}{150bp}{rgb(0bp)=(1,1,1);
rgb(45.80357142857143bp)=(1,1,1);
rgb(54.910714285714285bp)=(0.82,0.29,0.35);
rgb(62.5bp)=(0.73,0.15,0.22);
rgb(62.5bp)=(0.95,0.56,0.6);
rgb(100bp)=(0.95,0.56,0.6)}
\tikzset{every picture/.style={line width=0.75pt}} 


\draw    (73,20829) -- (119.02,20836.46) ;
\draw [shift={(98.58,20833.15)}, rotate = 189.21] [fill={rgb, 255:red, 0; green, 0; blue, 0 }  ][line width=0.08]  [draw opacity=0] (5.36,-2.57) -- (0,0) -- (5.36,2.57) -- cycle    ;
\draw    (369.82,20835.69) -- (416,20827) ;
\draw [shift={(395.47,20830.87)}, rotate = 169.34] [fill={rgb, 255:red, 0; green, 0; blue, 0 }  ][line width=0.08]  [draw opacity=0] (5.36,-2.57) -- (0,0) -- (5.36,2.57) -- cycle    ;
\draw    (317,20836) -- (369.82,20835.69) ;
\draw [shift={(346.01,20835.83)}, rotate = 179.67] [fill={rgb, 255:red, 0; green, 0; blue, 0 }  ][line width=0.08]  [draw opacity=0] (5.36,-2.57) -- (0,0) -- (5.36,2.57) -- cycle    ;
\draw    (236.63,20836.93) -- (282.61,20836.47) ;
\draw [shift={(262.22,20836.67)}, rotate = 179.43] [fill={rgb, 255:red, 0; green, 0; blue, 0 }  ][line width=0.08]  [draw opacity=0] (5.36,-2.57) -- (0,0) -- (5.36,2.57) -- cycle    ;
\draw    (119.02,20836.46) -- (177.82,20836.69) ;
\draw [shift={(151.02,20836.59)}, rotate = 180.23] [fill={rgb, 255:red, 0; green, 0; blue, 0 }  ][line width=0.08]  [draw opacity=0] (5.36,-2.57) -- (0,0) -- (5.36,2.57) -- cycle    ;
\draw    (177.82,20836.69) -- (236.63,20836.93) ;
\draw [shift={(209.83,20836.82)}, rotate = 180.23] [fill={rgb, 255:red, 0; green, 0; blue, 0 }  ][line width=0.08]  [draw opacity=0] (5.36,-2.57) -- (0,0) -- (5.36,2.57) -- cycle    ;
\draw    (118.78,20836.31) .. controls (120.45,20837.98) and (120.45,20839.64) .. (118.78,20841.31) .. controls (117.11,20842.98) and (117.11,20844.64) .. (118.78,20846.31) .. controls (120.45,20847.98) and (120.45,20849.64) .. (118.78,20851.31) .. controls (117.11,20852.98) and (117.11,20854.64) .. (118.78,20856.31) .. controls (120.45,20857.98) and (120.45,20859.64) .. (118.78,20861.31) .. controls (117.11,20862.98) and (117.11,20864.64) .. (118.78,20866.31) .. controls (120.45,20867.98) and (120.45,20869.64) .. (118.78,20871.31) .. controls (117.11,20872.98) and (117.11,20874.64) .. (118.78,20876.31) .. controls (120.45,20877.98) and (120.45,20879.64) .. (118.78,20881.31) -- (118.78,20881.31) ;
\draw    (130.6,20855.62) -- (130.6,20868.62) ;
\draw [shift={(130.6,20852.62)}, rotate = 90] [fill={rgb, 255:red, 0; green, 0; blue, 0 }  ][line width=0.08]  [draw opacity=0] (5.36,-2.57) -- (0,0) -- (5.36,2.57) -- cycle    ;
\draw    (177.82,20836.69) .. controls (179.48,20838.36) and (179.47,20840.03) .. (177.8,20841.69) .. controls (176.13,20843.35) and (176.12,20845.02) .. (177.77,20846.69) .. controls (179.42,20848.36) and (179.41,20850.03) .. (177.74,20851.69) .. controls (176.07,20853.35) and (176.06,20855.02) .. (177.71,20856.69) .. controls (179.37,20858.36) and (179.36,20860.03) .. (177.69,20861.69) .. controls (176.02,20863.35) and (176.01,20865.02) .. (177.66,20866.69) .. controls (179.31,20868.36) and (179.3,20870.03) .. (177.63,20871.69) .. controls (175.96,20873.35) and (175.95,20875.02) .. (177.6,20876.69) -- (177.58,20881.31) -- (177.58,20881.31) ;
\draw    (189.4,20855.62) -- (189.4,20868.62) ;
\draw [shift={(189.4,20852.62)}, rotate = 90] [fill={rgb, 255:red, 0; green, 0; blue, 0 }  ][line width=0.08]  [draw opacity=0] (5.36,-2.57) -- (0,0) -- (5.36,2.57) -- cycle    ;
\draw    (236.63,20836.93) .. controls (238.28,20838.6) and (238.27,20840.27) .. (236.6,20841.93) .. controls (234.93,20843.59) and (234.92,20845.26) .. (236.57,20846.93) .. controls (238.23,20848.6) and (238.22,20850.27) .. (236.55,20851.93) .. controls (234.88,20853.59) and (234.87,20855.26) .. (236.52,20856.93) .. controls (238.17,20858.6) and (238.16,20860.27) .. (236.49,20861.93) .. controls (234.82,20863.59) and (234.81,20865.26) .. (236.46,20866.93) .. controls (238.11,20868.6) and (238.1,20870.27) .. (236.43,20871.93) .. controls (234.76,20873.59) and (234.75,20875.26) .. (236.41,20876.93) .. controls (238.06,20878.6) and (238.05,20880.27) .. (236.38,20881.93) -- (236.38,20882.11) -- (236.38,20882.11) ;
\draw    (248.2,20856.42) -- (248.2,20869.42) ;
\draw [shift={(248.2,20853.42)}, rotate = 90] [fill={rgb, 255:red, 0; green, 0; blue, 0 }  ][line width=0.08]  [draw opacity=0] (5.36,-2.57) -- (0,0) -- (5.36,2.57) -- cycle    ;
\draw    (369.82,20835.69) .. controls (371.49,20837.35) and (371.5,20839.02) .. (369.84,20840.69) .. controls (368.18,20842.36) and (368.19,20844.03) .. (369.86,20845.69) .. controls (371.53,20847.36) and (371.53,20849.02) .. (369.87,20850.69) .. controls (368.21,20852.36) and (368.22,20854.03) .. (369.89,20855.69) .. controls (371.56,20857.35) and (371.57,20859.02) .. (369.91,20860.69) .. controls (368.25,20862.36) and (368.25,20864.02) .. (369.92,20865.69) .. controls (371.59,20867.35) and (371.6,20869.02) .. (369.94,20870.69) .. controls (368.28,20872.36) and (368.29,20874.03) .. (369.96,20875.69) .. controls (371.63,20877.35) and (371.64,20879.02) .. (369.98,20880.69) -- (369.98,20880.91) -- (369.98,20880.91) ;
\draw    (381.8,20855.22) -- (381.8,20868.22) ;
\draw [shift={(381.8,20852.22)}, rotate = 90] [fill={rgb, 255:red, 0; green, 0; blue, 0 }  ][line width=0.08]  [draw opacity=0] (5.36,-2.57) -- (0,0) -- (5.36,2.57) -- cycle    ;
\path  [shading=_o1zq98a2s,_anox4r0d0] (118.83,20872.29) .. controls (124.05,20872.32) and (128.25,20876.38) .. (128.22,20881.36) .. controls (128.19,20886.34) and (123.94,20890.36) .. (118.72,20890.33) .. controls (113.51,20890.31) and (109.3,20886.25) .. (109.33,20881.26) .. controls (109.37,20876.28) and (113.62,20872.27) .. (118.83,20872.29) -- cycle ; 
 \draw   (118.83,20872.29) .. controls (124.05,20872.32) and (128.25,20876.38) .. (128.22,20881.36) .. controls (128.19,20886.34) and (123.94,20890.36) .. (118.72,20890.33) .. controls (113.51,20890.31) and (109.3,20886.25) .. (109.33,20881.26) .. controls (109.37,20876.28) and (113.62,20872.27) .. (118.83,20872.29) -- cycle ; 

\path  [shading=_ol8dd6hbu,_19cp6sbk5] (177.63,20872.29) .. controls (182.85,20872.32) and (187.05,20876.38) .. (187.02,20881.36) .. controls (186.99,20886.34) and (182.74,20890.36) .. (177.52,20890.33) .. controls (172.31,20890.31) and (168.1,20886.25) .. (168.13,20881.26) .. controls (168.17,20876.28) and (172.42,20872.27) .. (177.63,20872.29) -- cycle ; 
 \draw   (177.63,20872.29) .. controls (182.85,20872.32) and (187.05,20876.38) .. (187.02,20881.36) .. controls (186.99,20886.34) and (182.74,20890.36) .. (177.52,20890.33) .. controls (172.31,20890.31) and (168.1,20886.25) .. (168.13,20881.26) .. controls (168.17,20876.28) and (172.42,20872.27) .. (177.63,20872.29) -- cycle ; 

\path  [shading=_5fkrmtb6k,_qnr8f0d09] (236.43,20873.09) .. controls (241.65,20873.12) and (245.85,20877.18) .. (245.82,20882.16) .. controls (245.79,20887.14) and (241.54,20891.16) .. (236.32,20891.13) .. controls (231.11,20891.11) and (226.9,20887.05) .. (226.93,20882.06) .. controls (226.97,20877.08) and (231.22,20873.07) .. (236.43,20873.09) -- cycle ; 
 \draw   (236.43,20873.09) .. controls (241.65,20873.12) and (245.85,20877.18) .. (245.82,20882.16) .. controls (245.79,20887.14) and (241.54,20891.16) .. (236.32,20891.13) .. controls (231.11,20891.11) and (226.9,20887.05) .. (226.93,20882.06) .. controls (226.97,20877.08) and (231.22,20873.07) .. (236.43,20873.09) -- cycle ; 

\path  [shading=_lmmfwi4xg,_ab148gheg] (370.03,20871.89) .. controls (375.25,20871.92) and (379.45,20875.98) .. (379.42,20880.96) .. controls (379.39,20885.94) and (375.14,20889.96) .. (369.92,20889.93) .. controls (364.71,20889.91) and (360.5,20885.85) .. (360.53,20880.86) .. controls (360.57,20875.88) and (364.82,20871.87) .. (370.03,20871.89) -- cycle ; 
 \draw   (370.03,20871.89) .. controls (375.25,20871.92) and (379.45,20875.98) .. (379.42,20880.96) .. controls (379.39,20885.94) and (375.14,20889.96) .. (369.92,20889.93) .. controls (364.71,20889.91) and (360.5,20885.85) .. (360.53,20880.86) .. controls (360.57,20875.88) and (364.82,20871.87) .. (370.03,20871.89) -- cycle ; 

\draw (454,20853) node [anchor=north west][inner sep=0.75pt]    {$=i\mathcal{A}_L (p\to p'),$};
\draw (58,20824) node [anchor=north west][inner sep=0.75pt]    {$p$};
\draw (421,20819) node [anchor=north west][inner sep=0.75pt]    {$p'$};
\draw (289,20832) node [anchor=north west][inner sep=0.75pt]    {$\cdots $};
\draw (135.6,20854.62) node [anchor=north west][inner sep=0.75pt]    {$\ell _{1}$};
\draw (194.4,20854.62) node [anchor=north west][inner sep=0.75pt]    {$\ell _{2}$};
\draw (253.2,20855.42) node [anchor=north west][inner sep=0.75pt]    {$\ell _{3}$};
\draw (385.8,20856.22) node [anchor=north west][inner sep=0.75pt]    {$\ell _{L+1}$};

\end{tikzpicture}
\]
whose explicit expression in the leading geometric-optics limit is
\[ \label{sclassL}
    i\mathcal{A}_L(p &-q \to p)=i^L\int \hat{\dd}^4\ell_1\cdots\hat{\dd}^4\ell_{L+1}\,\hat{\delta}^4(\ell_{12\cdots L+1}-q)
    \\&
    \hspace{3cm}
    \times
    \frac{-2iQ \tilde{A}(\ell_1)\cdot p}{2p\cdot\ell_1+i\epsilon} \frac{-2iQ \tilde{A}(\ell_2)\cdot p}{2p\cdot\ell_{12}+i\epsilon}\cdots \frac{-2iQ \tilde{A}(\ell_{L+1})\cdot p}{2p\cdot\ell_{12\cdots L}+i\epsilon},
\]
defining $\ell_{i\cdots j}=\sum_{k=i}^j \ell_k$.
Note again that due to the linearity of the Kerr-Schild interaction Lagrangian there are no contact terms here. The next step is to show that equation \eqref{sclassL} is a pure convolution (product) in momentum (position) space of the tree-level result \eqref{treelvl}.  This is immediate  in the leading geometric-optics limit since the dependence of each interaction insertion $\tilde{A}(\ell_i)\cdot p$ is only through the Fourier phase $e^{ix_i\cdot \ell_i}$,  just like in equation \eqref{leadinga0}. This fact, together with the use of the eikonal identity that turns linearized propagators into delta functions 
\cite{Akhoury:2013yua}
\[
\sum_\sigma \frac{i^L\,\hat{\delta}(p \cdot \ell_{1\cdots L+1})}{(p \cdot \ell_{\sigma(1)} +i \epsilon) \cdots ( p \cdot \ell_{\sigma(1) \cdots \sigma(L)}+ i\epsilon)} = \prod_{i=1}^{L+1} \hat{\delta}(p \cdot \ell_i),
\]
allows us to resum all the leading-in-$\eta$ loops in equation \eqref{sclassL} with a factor of $1/L!$.  The sum over $\sigma$ runs over permutations of the loop variables.
Then, in position space one finds
\[
 \int \hat{\dd}^4 q\, \hat{\delta}(2 p\cdot q)i\mathcal{A}_{L}(q)e^{-iq\cdot b(v)}=  \frac{1}{(L+1)!}
    \left[\int \hat{\dd}^4 \ell\, \hat{\delta}(2 p\cdot \ell)i\mathcal{A}^{(0)}_{0}(\ell)e^{-i\ell\cdot b(v)}\right]^{L+1},
\]
where $i\mathcal{A}^{(0)}_{0}(\ell)$ is the leading-in-$\eta$ tree-level result. Finally, we end up with the desired result
\begin{align}\label{eq:outgoingExp}
 \braket{p|S|\psi}
    &= \int \dd v \,\varphi(v) e^{i p\cdot b(v)} \
    \left[1+\sum_{L=0}^\infty\int \hat{\dd}^4 q\, \hat{\delta}(2 p\cdot q)i\mathcal{A}_{L}(p-q\to p)e^{-iq\cdot b(v)}\right]\nonumber\\&=
    \int \dd v \,\varphi(v) e^{i p\cdot b(v)}\exp\left[2i\alpha\log(-v/\mu)\right],
\end{align}
having introduced a ``fine structure constant" 
\[
\alpha \equiv  \frac{QQ_\text{B}}{4\pi}.
\]

We end this section noting that the exponentiation argument given above essentially proceeded in the same way as it did in gravity~\cite{Aoude:2024sve}. The reason for this is simple. In the spirit of the classical double copy of \cite{Monteiro:2014cda} only tensor numerators differ between the two theories:\footnote{Actually, this also applies to the BCJ double copy of \cite{Bern:2008qj}.}
\[\label{dcks}
\text{GR}: \,\, \Theta(t+r)\frac{\sfk^\mu \sfk^\nu}{r} \quad \leftrightarrow  \quad \text{EM}:  \,\, \Theta(t+r)\frac{\sfk^\mu }{r} \,.
\]
Denominators are in common between the two theories. Going from gravity to electrodynamics (and vice versa) through this prescription results in leading $L$-loop  amplitude numerators (see for instance   eq. \eqref{sclassL} above) which   depend on transfer momenta in a trivial manner, regardless of the theory. In fact, this is all that is needed to exponentiate the leading eikonal phase according to the discussion below eq. \eqref{sclassL}.

\subsection{The $\sqrt{\text{Hawking}}$ amplitude and Bogoliubov coefficients}

At this point we are in position to compute the Bogoliubov coefficient that will determine the spectrum of the massless scalar. To this end, we find it useful to first define the $1\to 1$ $\sqrt{\text{Hawking}}$ amplitude  by
\[\label{sqrthamp}
\mathcal{A}(E) \equiv  \int \dd v \,\varphi(v) e^{i p\cdot b(v)}\exp\left[2i\alpha\log(-v/\mu)\right].
\] 
To proceed we must now specify the details of the initial wavepacket.
A choice of $\varphi(v)$ localised at a specific advanced time (say $v_0$) would allow us to track the time dependence of the final-state radiation.
Here we will not examine this dependence in detail, and so we instead choose a simple spherically symmetric wavepacket, namely a $\ell=0 $ spherical harmonic for the incoming state 
\[
|\psi\rangle= \int  \frac{\dd \Omega_{{{p}}} }{4 \pi}\ket{E_0, E_0\, \hat{\vec{p}}}\,\,\, \Leftrightarrow  \,\,\, \varphi(v) = \frac{2 \pi}{E_0}\, e^{-iE_0v} \,,
\]
yielding the following Bogoliubov coefficients:
\[
\mathcal{A}(E,E_0)=  \frac{2\pi}{E_0} \int \dd v \,  e^{i v(E-E_0)}\exp\left[2i\alpha\log(-v/\mu)\right],
\]
\[
\mathcal{B}(E,E_0)=  \frac{2\pi}{E_0} \int \dd v \,  e^{i v(E+E_0)}\exp\left[2i\alpha\log(-v/\mu)\right]=\mathcal{A}(E,-E_0).
\]
According to \cite{Aoude:2024sve} these are simply obtained from the $\sqrt{\text{Hawking}}$ amplitude \eqref{sqrthamp} and by its crossed version, multiplying by a kinematic factor.

As discussed in section~\ref{sec:resum}, the position-space wavefunction for the scattering particle has support only for $v<0$ if it starts inside the infalling spherical charge shell. With this in mind we restrict the integral to only negative values, which conveniently avoids the branch cut of the logarithm. The integrals can now be performed explicitly, with the results 
\[\label{eq:bogAfinal}
\mathcal{A}(E,E_0)&=  \frac{2\pi}{E_0} \int_{-\infty}^0 \dd v \, e^{i v(E-E_0)}\exp\left[2i\alpha\log(-v/\mu)\right]
\\&=
\frac{2\pi}{E_0}  (i (E-E_0) )^{-1-2i \alpha}\,\Gamma \left(1+2i \alpha\right),
\]
and
\[\label{eq:bogB}
\mathcal{B}(E,E_0)&=
\frac{2\pi}{E_0}  (i (E+{E_0}) )^{-1-2i \alpha}\,\Gamma \left(1+2i \alpha\right).
\]
It is common in the literature on Hawking radiation to expand the energy factors in equation~\eqref{eq:bogAfinal} in the region $E \ll E_0$. 
However, in this electromagnetic case, the force may be either attractive or repulsive. 
Since in the repulsive case the final energy is in fact the larger, we do not perform this expansion here to avoid making unnecessary assumptions.

\section{Bogoliubov coefficients via semiclassical ray-tracing}
\label{sec:Bogray}

In the last subsection, we saw how to determine Bogoliubov coefficients using Feynman diagrams and the double copy.
Here we confirm our understanding of the situation by reproducing these Bogoliubov coefficients using Hawking's original approach, based on an understanding of the trajectories of null rays in the background.

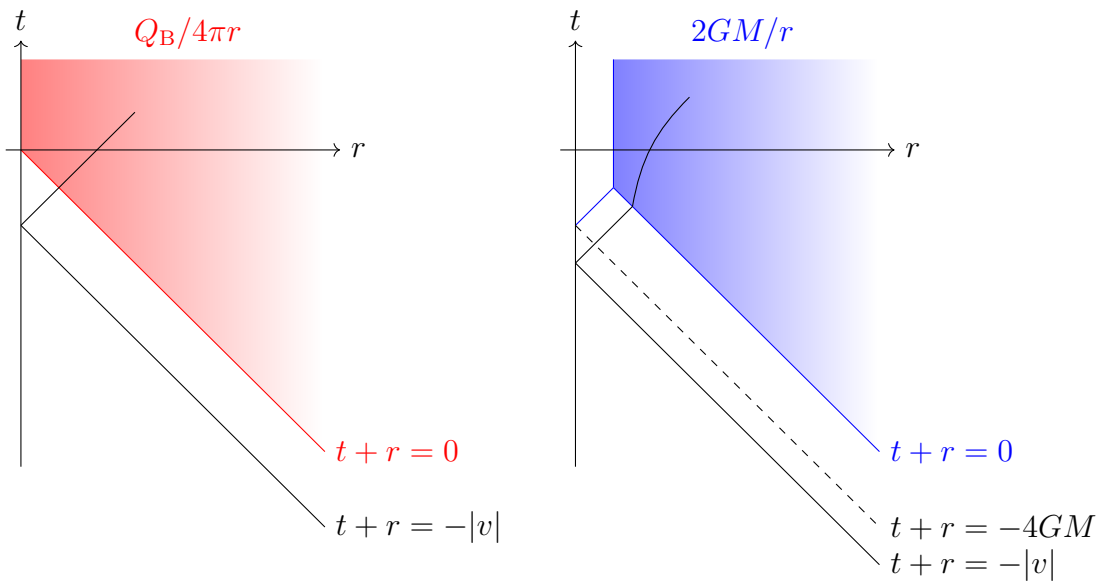
\begin{figure}[!htb]
\begin{subfigure}[!t]{.5\textwidth}
\centering
\begin{tikzpicture}[domain=0:4]
  \path [shading = axis,rectangle, left color=red!50, right color=white,shading angle=90, anchor=left] (0,0) -- (0,1.2) -- (4.0,1.2) -- (4.0,-4.0);

  \draw[color=red] plot (\x,{-\x}) node[right] {$t+r=0$};
  \draw[red] (2.2,1.2) node[above] {$Q_{\rm B}/4\pi r$};
  
  \draw[color=black] plot (\x,{-\x-1}) node[right] {$t+r=-|v|$};
  
  \draw[-] (0,-1.0) -- (1.50,-1+1.5);

  \draw[->] (-0.2,0) -- (4.2,0) node[right] {$r$};
  \draw[->] (0,-4.2) -- (0,1.45) node[above] {$t$};
\end{tikzpicture}
\end{subfigure}%
\begin{subfigure}[!t]{.5\textwidth}
\centering
\vspace{0.5cm}
\begin{tikzpicture}[domain=0:4]
  \path [shading = axis,rectangle, left color=blue!50, right color=white,shading angle=90, anchor=left] (0.50,-1+0.5) -- (0.50,1.2) -- (4.0,1.2) -- (4.0,-4.0);

  \draw[color=blue, domain= 0.5:4.0] plot (\x,{-\x}) node[right] {$t+r=0$};
  \draw[blue] (2.2,1.2) node[above] {$2GM/r$};
  
  \draw[color=black,dashed] plot (\x,{-\x-1}) node[right] {$t+r=-4GM$};
  \draw[color=blue] (0,-1.0) -- (0.50,-1+0.5);
  \draw[color=blue] (0.50,-1+0.5) -- (0.50,1.2);

  \draw[color=black] plot (\x,{-\x-1.5}) node[right] {$t+r=-|v|$};
  \draw[color=black] (0,-1.5) -- (0.75,-0.75);
  
  \draw [black, out=80,in=225] (0.75,-0.75) to (1.5,0.70);

  \draw[->] (-0.2,0) -- (4.2,0) node[right] {$r$};
  \draw[->] (0,-4.2) -- (0,1.45) node[above] {$t$};
\end{tikzpicture}
\end{subfigure}%
\caption{Scattering on a $\sqrt{\rm Vaidya}$ (Vaidya) background on the left (right) figure. In the $\sqrt{\rm Vaidya}$ case, the infalling shell of radiation creates a charge at $t+r=0$, we set  our initial state to be inside this infalling shell (unshaded region). For the Vaidya case, the infalling shell of radiation creates a black hole  at $t+r=0$ (shaded region). In \cite{Aoude:2024sve}, the authors considered initial states with $t+r<-4GM$ that do not fall inside the horizon.}
\label{Fig:SpacetimeDiagrams}
\end{figure}

The trajectories of interest to us are those of massless probe charges interacting with the \rvaidya background.
These interactions are attractive (as in the gravitational case) provided the product of the background charge $Q_\text{B}$ and probe charge $Q$ is negative.
We take the action for the probe to be
\[
\label{eq:actionDef}
I = \int \dd \lambda \left[ -\frac12 \left( \frac{\dd x}{\dd \lambda}\right)^2 - Q A_\mu(x(\lambda)) \frac{\dd x^\mu}{\dd \lambda} \right] \,,
\]
and assume that the probe falls radially towards the origin starting from some fixed advanced time $v$ on $\scri^-$.
The basic idea is that the probe particle passes through the origin before the shell of massless radiation (see Fig.~\ref{Fig:SpacetimeDiagrams}).
Since the shell falls along the line $t + r = 0$, we must take $v < 0$ for our probe to pass through the origin first.

In this electromagnetic case, the spacetime trajectory of the probe is completely trivial because it is constrained to move along a lightcone in flat space.
Thus, the incoming portion of the probe trajectory is the straight line $t+r = v$.
On the outgoing part of the trajectory, instead $t -r$ is constant; continuity at the origin then implies that $t-r = v$ on the outgoing part of the trajectory. We may choose the parameter $\lambda$ on the particle trajectory to be $-r$ in the incoming part of the trajectory and $+r$ on the outgoing part.

Nevertheless as the particle passes into the non-trivial Coulomb field it accumulates a phase $e^{i I_\textrm{int}}$ given by integrating the interaction action along its worldline.
For radial motion, the interaction action in equation~\eqref{eq:actionDef} simplifies to
\[
I_\textrm{int} = \int \dd r \left[ - \frac{Q Q_\text{B}}{4\pi r} \Theta(t(r)+r) \frac{\dd}{\dd r}(t(r)+r) \right] \,,
\]
taking the parameter $\lambda$ to be the radius on the outgoing part of the trajectory.
In this region, $t(r) = v + r$, so the theta function requires $r > -v /2$.
The interaction action is
\[\label{phasesqrt}
I_\textrm{int} &= \int_{-v/2}^\infty \dd r \left[ - \frac{Q Q_\text{B}}{2\pi r} \right]= 2 \alpha \log(-v/\mu) \,.
\]
Here we dropped the IR phase, consistent with equation~\eqref{eq:IRphase}.
As a result, the wavepacket of the outgoing particle involves the phase factor
\[
e^{iI_\textrm{int}} = \exp [2 i \alpha  \log(-v/\mu)] \,,
\]
consistent with equation~\eqref{eq:outgoingExp}.

\section{Number distribution}
\label{sec:number}

Given the $\sqrt{\text{Hawking}}$ amplitude, we can compute the differential number spectrum. This is defined as the integral of the generalized amplitude $\bra{\Omega} S^\dagger a^\dagger a S\ket{\Omega}$, \text{i.e}
\begin{align}
 n = \int\! \dd\Phi(p) \bra{\Omega}S^\dagger a^\dagger(p) a(p) S\ket{\Omega}
 = \int\! \dd\Phi(p,k) B^*(p,k)B(p,k)
\end{align}
where the Bogoliubov coefficient $B(p,k)$ was derived in~\cite{Aoude:2024sve} as a generalised amplitude~\cite{Schwarz:2019ggp,Schwarz:2019npn,Caron-Huot:2023ikn,Caron-Huot:2023vxl}. $B(p,k)$ and $A(p,k)$ are typically related to the Bogoliubov transformation. Both are defined as
\begin{align}
    B(p,k) = \bra{\Omega} a(k) S^\dagger a(p) S \ket{\Omega},
    \qquad
    A(p,k) = \bra{\Omega} S^\dagger a(p) S a (k) \ket{\Omega}
\end{align}
Relating $B(p,k)$ and $A(p,k)$ as in~\cite{Aoude:2024sve}, the number operator is given in terms of amplitude squared by
\begin{align}
    \dd n = \frac{E\dd E}{4\pi^2}\frac{E_0\dd E_0}{4\pi^2} \frac{E_0^2}{E^2}\,\, |{\cal B}(E)|^2 \,.
\end{align}
The $\sqrt{\text{Hawking}}$ pair production generalised amplitude ${\cal B}(E)$ after integration as described in Section~\ref{sec:BogQFT} is
\[
\mathcal{B}(E,E_0)&=
\frac{2\pi}{E_0}  (i (E+{E_0}) )^{-1-2i \alpha}\,\Gamma \left(1+2i \alpha\right).
\]
where we have defined $\alpha \equiv QQ_{\text{B}}/(4\pi)$ to resemble the QED fine-structure constant. Squaring the absolute value of this amplitude and using Euler’s reflection formula,\footnote{c.f. Eq.\,4.53 in~\cite{Aoude:2024sve}.} we obtain
\begin{align}\label{bogoliubovb}
	|\mathcal{B}(E)|^2 
	&= \frac{(2\pi)^2}{E^2_0} \frac{e^{2 \pi \alpha}}{(E_0+E)^2}|\Gamma(1+2i\alpha)|^2
	=\frac{(2\pi)^2}{E^2_0(E+E_0)^2}  
    \frac{(-4\pi\alpha)}{e^{-4\pi \alpha}-1} \geq 0\,.
\end{align}
Finally, the differential number operator is
\begin{align}\label{spec}
    \dd n &= \dd E \,\dd E_0  \frac{E_0}{E(E+E_0)^2}\frac{-\alpha}{\pi}
    \frac{1}{e^{-4\pi \alpha}-1} + \cdots,
\end{align}
which is positive regardless of the sign of $\alpha$. Here, the dots indicate contributions from diagrams which are higher order in $\eta$ (see equation~\eqref{golimit}) which we have not included.

There are crucial differences compared to the gravitational case, whose number distribution takes the following form (c.f. Eq (4.54) ~\cite{Aoude:2024sve}):
\begin{equation}
\dd n=\dd E \dd E_0\frac{2GM}{\pi E_0}\frac{1}{e^{8\pi GME}-1}.
\label{dngrav}
\end{equation}
In the electromagnetic case, the force after crossing can now be both attractive and repulsive.  Let us further elaborate  on this important point, taking for definiteness $Q_{\text{B}} >0$. For clarity we can  define the numbers of the two distinct processes depending on the sign of the charge $Q$:
\[
    \dd n_\pm &= \dd E \,\dd E_0  \frac{E_0}{E(E+E_0)^2}\frac{\pm|Q Q_{\text{B}}|}{4\pi^2}
    \frac{1}{e^{\pm |QQ_{\text{B}}|}-1} \,.
\] 
First we consider the case $Q\!<\!0$. Then, crossing and  charge conservation imply that the created pair is composed of one positive particle falling radially towards the positive background, and a negative one escaping to infinity: the emission is described by $\dd n_+$ and it is suppressed when $|Q Q_{\text{B}}|$ grows. Conversely, if $Q\!>\!0$ the infalling particle is negatively charged and is thus attracted by the heavy source. In this case the spectrum $\dd n_-$ will be enhanced and therefore dominates the final radiation. 
We learn that electrodynamics naturally gives rise to two physically distinct processes, both of which can be considered to be single copies of Hawking radiation.

The energy independence in the obtained distribution is another important point. In fact, the exponential $e^{-4\pi \alpha}$ now depends only on the strength and sign of the interaction, but not on the energy of the probe. In turns, this leads to a spectrum that seems non-thermal, at least in the na\"{i}ve sense. We will try to interpret this result soon, but the reason why the $\sqrt{\text{Vaidya}}$ phase \eqref{phasesqrt} is dimensionless is clear from the double copy point of view: the gravitational coupling is dimensionful, unlike the EM one. Then, the double copy replacement of equation \eqref{eq:SqrtVaidya} maps  a dimensionful quantity into a number or, to put it differently, naturally connects  two dimensionless physical quantities together
\[
2GME \quad \leftrightarrow \quad -\alpha.
\label{DCreplace1}
\]
This is one way to understand how $\sqrt{\rm Vaidya}$ yields trivial energy dependence for $\dd n$. However, known results about the double copy allow us to interpret things further. We may understand eq.~(\ref{DCreplace1}) in more detail by reinstating the gauge theory coupling constant accompanying each charge:
\begin{equation}
  Q\rightarrow gQ,\quad Q_{\text{B}}\rightarrow g Q_{\text{B}}.
\label{Qrenorm}
\end{equation}
We may then understand eq.~(\ref{DCreplace1}) as the sequential set of replacements
\begin{equation}
  \frac{\kappa}{2}\rightarrow g,\quad Q_{\text{B}}\rightarrow M,\quad
  -Q\rightarrow E,
  \label{DCreplace2}
\end{equation}
where $\kappa^2=\sqrt{32\pi G}$ is the conventional gravitational coupling in terms of Newton's constant. The first replacement in eq.~(\ref{DCreplace2}) is the usual replacement of coupling constants between gauge theory and gravity that occurs in the original Kerr-Schild double copy of ref.~\cite{Monteiro:2014cda}, as well as the BCJ double copy for scattering amplitudes~\cite{Bern:2010ue,Bern:2010yg}. The second and third replacements correspond to the systematic replacement of kinematic information in the gravity theory, with charge information in the gauge theory. 
Since the double copy strips off colour information, either sign of charge can be obtained in the single copy and if we want to focus on the attractive case, this fixes $-Q = |Q| \to E$.
That one involves a mass and the other an energy replacement reflects the static nature of the overall mass $M$ once the outgoing particle has crossed the horizon, and the dynamic nature of the outgoing particle with energy $E$. That the replacements of eq.~(\ref{DCreplace2}) are correct is commensurate with standard lore on the classical double copy. What is interesting, however, is the question of whether the single copy number spectrum has any kind of thermal interpretation, given its strikingly different energy dependence compared with the gravitational case.
To examine this further, we note that a more general thermal distribution one can consider is as follows:
\begin{align}
	\bar{n}(\varepsilon) = \frac{1}{e^{(\varepsilon - \mu_{\rm c})/k_{\rm B}T}-1},
    \label{num2}
\end{align}
where $\varepsilon$ is the energy of the system, and $\mu_c$ the chemical potential, where the latter is thermodynamically conjugate to particle number. Comparison with eq.~(\ref{spec}) reveals that the single copy thermal spectrum behaves as if the energy term in eq.~(\ref{num2}) is absent, and one instead has a pure chemical potential dependence. Indeed, there are good physical reasons why the single copy should produce such a spectrum. In gravity, one may count the total energy $E$ of the system by summing up the energies of each individual particle emitted from the black hole. Upon taking the single copy, the energy is replaced by the charge, such that the analogue of counting up particle energies is counting up their individual charges. However, charge is a proxy for particle number, such that one expects the single copy to take the energy term in a thermal distribution, and replace it with the chemical potential term. 

So much for how the chemical potential arises. However, there remains the question of how physical a thermal distribution is, which has a chemical potential term but no energy dependence. One way to understand such a distribution is that it can be obtained from eq.~(\ref{num2}) by taking a limit of high temperature ($T\rightarrow\infty$), but with $\beta\mu_{\rm c}$ held fixed. Credence for this interpretation can be obtained by considering Einstein-Maxwell solutions. In previous cases of the classical double copy, the Kerr-Schild single copy of a given (pure) gravity solution can be understood by taking the $G\rightarrow0$ limit of a corresponding Einstein-Maxwell solution, such that the single copy gauge field can be viewed as the gauge field of the Einstein-Maxwell solution, with gravity turned off. The thermal spectrum of the Reissner-Nordstrom black hole (a charged analogue of Schwarzschild, albeit with electric potential chosen to be zero at the horizon) has both an energy and a chemical potential term, as in eq.~(\ref{num2}). If one then takes $G\rightarrow 0$, then $\beta\rightarrow 0$, such that the energy term in the thermal spectrum vanishes, but the chemical potential would survive if $\beta\mu_{\rm c}$ is fixed. We further note that the combination $e^{\beta\mu_{\rm c}}$ is known in the statistical physics literature as the \textit{fugacity}. Here, it plays the special role of the quantity picked out by making the replacements of eq.~(\ref{DCreplace2}) in the number spectrum of Hawking radiation. This may provide clues towards a more rigorous thermal interpretation of our somewhat speculative comments presented here.

\section{Conclusions}
\label{sec:conclude}

The connection between gravity and gauge theory through the double copy continues to yield significant results for both quantum field theory and classical solutions. In this work, we extended this analysis to the case of Hawking radiation.  We realized that this can be done by considering scattering problems on time-dependent backgrounds. Then, our study starts from the Vaidya solution~\cite{Vaidya:1966zza}: a non-static solution of the Einstein field equations describing the birth of a black hole from an infalling shell of radiation. Here, our analysis shows that the non-static mass dependence, $M(t+r)$, of the Vaidya background naturally defines a source of dynamical charge distribution  $Q_{\rm B}(t+r)$ which we call $\sqrt{\text{Vaidya}}$. We implemented this correspondence using the classical double copy framework of \cite{Monteiro:2014cda}, this is  made easier by the known existence of Kerr-Schild coordinates of the Vaidya metric.

From here on, we can consider scattering on the electromagnetic background as usual. We scatter a light charged particle with the  $\sqrt{\text{Vaidya}}$ dynamical source and compute the eikonal function by resumming ladder diagrams \cite{DiVecchia:2023frv, Cristofoli:2021jas}. Our probe is taken to be massless for convenience: one can  simply imagine it being much lighter than the background.
The Feynman diagrams and their structure are extremely similar to the gravity ones of \cite{Aoude:2024sve}, this is a known consequence of the double copy. In fact, the difference between the two sides of the duality is only in the numerators -- which are squared in GR -- whereas propagators are unchanged and equal. This signals the fact that both theories fall off with a $1/r$ behavior. All of this translates into a similar resummation pattern of the one in~\cite{Aoude:2024sve}, found using amplitudes for the first time. However, there is one crucial difference: we find an energy-independent eikonal. The single copy of the Hawking eikonal phase $4GME\log(-v)$ translates into a dimensionless prefactor of the known leading logarithm $2\alpha \log(-v)$.
Correspondingly, the thermodynamic distribution we encounter is characterised by its fugacity rather than a temperature.
This is further confirmed by considerations of the double copy and the charged Reissner-Nordstr\"om black hole, which is known to have a Hawking radiation distribution with both a temperature and a chemical potential, where the latter arises from the non-zero gauge field~\cite{Hawking:1975vcx,Iso:2006wa}. 

We can easily identify future directions to take. One obvious task is to explore higher orders in perturbation theory, or to compute NLO eikonal contributions. It was shown in \cite{Aoude:2024sve} that these resum into horizon contributions in gravity but we know there is no EM horizon in the $\sqrt{\text{Vaidya}}$ case. Perhaps here the framework of \cite{Chawla:2023bsu} could become useful to interpret the horizons. Another direction involves characterising the EM scattering without the use of a background. One way to go about this is to model the infalling shell of radiation with coherent states and to then compute full QFT amplitudes of the probe interacting with the coherent state modes. Another option is to endow the source with additional degrees of freedom, such as spin. In the EM case, this is known as the $\sqrt{\text{Kerr}}$ solution \cite{Arkani-Hamed:2019ymq}, but a time-dependent analogue has not been defined yet.  We leave these exciting research avenues to future investigations.

\acknowledgments
We thank Karthik Rajeev for valuable discussions.
We thank Anton Ilderton, William Lindved and Karthik Rajeev for sharing a draft of their paper~\cite{Ilderton:2025} prior to publication and for coordinating the submission. R.A. is supported by UK Research and Innovation (UKRI) under the UK government’s Horizon
Europe Marie Sklodowska Curie funding guarantee grant [EP/Z000947/1].
D.O.C is supported by the European Research Council under Advanced Investigator grant ERC–AdG–101200505 and by the STFC grant “Particle Theory at the Higgs Centre”.
M.S. has been supported by the European Research Council under Advanced Investigator grant
ERC–AdG–885414, by the U.S. Department of Energy (DOE) under award number DE-SC0009937, and by the European Research Council (ERC) Horizon Synergy Grant “Making Sense of the Unexpected in the GravitationalWave Sky” grant agreement no. GWSky–101167314. M.S. also acknowledges support from the Mani L. Bhaumik institute for Theoretical Physics. CDW is supported by the UK Science and Technology Facilities Council (STFC) Consolidated Grant ST/P000754/1 “String theory, gauge theory and duality”.

\appendix

\section{Charge conservation and balance equation}\label{appendix}
In the section we will show that the \emph{total} time dependent $\sqrt{\text{Vaidya}}$ source considered in this article obeys current conservation.

The total current of the source is constituted by two contributions: the radiative part \eqref{current} and the Coulombic component
\[
j^\mu_\text{Coul.}=   Q \Theta(t+r) \delta^{(3)}(\Vec{x}) u^\mu,
\] 
which is also time-dependent.
It is an essential requirement of our setup's consistency to verify that the total current
\[
J^\mu = j^\mu_\text{Coul.}+j^\mu= Q \Theta(t+r) \delta^{(3)}(\Vec{x}) u^\mu+\frac{Q}{4\pi r^2} \delta(t+r) \, \sfk^\mu,
\]
is indeed conserved. 
Let's first compute the Coulombic contribution, writing 
\[
\sfk^\mu=u^\mu-n^\mu \,\,\,\text{with}\,\,\, u^2=1,\,\,\,n^2=-1,\,\,\,n\cdot u=0, \,\,\, n\cdot \sfk=1=u\cdot \sfk.
\]
We get
\[
\partial_\mu j^\mu_\text{Coul.}&=Q\delta(t+r)\delta^{(3)}(\vec x)\sfk\cdot u+Q \Theta(t+r) \delta^{'(3)}(\Vec{x}) u\cdot n\\&=Q\delta(t+r)\delta^{(3)}(\vec x).
\]
Moving onto the radiative part we have to compute
\[
\partial_\mu j^\mu =
\frac{Q}{4\pi} \delta(t+r) \,\partial_\mu \left( \frac{\sfk^\mu}{r^2}\right),
\]
where we simplified using $\sfk^2=0$. To obtain the correct distribution from this term we need to treat the $r\to 0 $ limit more carefully. One way to do this, following Jackson \cite{Jackson:1998nia}, is with a regulator $\sigma$
\[
r=|\vec x|\to  \sqrt{\vec x ^2+\sigma^2},
\]
which is sent to zero eventually. One finds  
\[
\partial_\mu \left( \frac{\sfk^\mu}{r^2}\right)=\frac{-3\sigma^2}{(\vec x^2+\sigma^2)^{5/2}}.
\]
It is easy to see that this distribution is a spatial delta function by integrating against a test function:
\[
\lim_{\sigma\to 0}\int \dd^3 \vec x f(\vec x)\frac{-3\sigma^2}{(\vec x^2+\sigma^2)^{5/2}}=-3\lim_{\sigma\to 0}\int \dd^3 \vec y f(\sigma \vec y)\frac{1}{(\vec y^2+1)^{5/2}}=-4\pi f(\vec 0),
\]
having rescaled the integration variable  as $\vec x\to \sigma \vec y$.
Thus, in a distributional sense 
\[
\partial_\mu \left( \frac{\sfk^\mu}{r^2}\right)=-4\pi \delta^{(3)}(\vec x),
\]
which ensures that total energy-momentum is conserved  for the non-static $\sqrt{\text{Vaidya}}$ source since the two contributions balance one another in a non trivial manner
\[
\partial_\mu J^\mu=Q\delta(t+r)\delta^{(3)}(\vec x)+\frac{Q}{4\pi}\delta(t+r)(-4\pi \delta^{(3)}(\vec x)) =0.
\]

\bibliographystyle{JHEP}
\bibliography{SqrtHawking.bib}
\end{document}